\documentclass[sigconf, anonymous=false,natbib=false]{acmart}
\usepackage{subcaption}

\RequirePackage[
  datamodel=acmdatamodel,
  style=acmnumeric,
  ]{biblatex}

\setcopyright{acmlicensed}
\copyrightyear{2025}
\acmYear{2025}
\acmDOI{XXXXXXX.XXXXXXX}

\acmCodeLink{https://git.isir.upmc.fr/mat_vast/sigir25-matching-signals}
\acmConference[WExIR25]{Workshop on Explainability in Information Retrieval at SIGIR 2025}{July 17, 2025}{Padova, Italy}

\begin{document}

\title{Understanding Matching Mechanisms in Cross-Encoders}

\author{Mathias Vast}
\orcid{0009-0007-4612-717X}
\affiliation{%
  \institution{Sinequa}
    \city{Paris}
  \country{France}
}
\affiliation{%
  \institution{Sorbonne Université, CNRS, Institut des Systèmes
Intelligents et de Robotique,}
  \city{Paris}
  \country{France}
}
\email{mathias.vast@isir.upmc.fr}

\author{Basile Van Cooten}
\orcid{0009-0002-0234-917X}
\affiliation{%
  \institution{Sinequa}
  \city{Paris}
  \country{France}
}
\email{vancooten@sinequa.com}

\author{Laure Soulier}
\orcid{0000-0001-9827-7400}
\affiliation{%
  \institution{Sorbonne Université, CNRS, Institut des Systèmes
Intelligents et de Robotique,}
  \city{Paris}
  \country{France}
}
\email{laure.soulier@isir.upmc.fr}

\author{Benjamin Piwowarski}
\orcid{0000-0001-6792-3262}
\affiliation{%
  \institution{Sorbonne Université, CNRS, Institut des Systèmes
Intelligents et de Robotique,}
  \city{Paris}
  \country{France}
}
\email{benjamin.piwowarski@isir.upmc.fr}

\renewcommand{\shortauthors}{Vast et al.}
\acmArticleType{Review}

\begin{abstract}
Neural IR architectures, particularly cross-encoders, are highly effective models whose internal mechanisms are mostly unknown. Most works trying to explain their behavior focused on high-level processes (e.g., what in the input influences the prediction, does the model adhere to known IR axioms) but fall short of describing the matching process. Instead of Mechanistic Interpretability approaches which specifically aim at explaining the hidden mechanisms of neural models, we demonstrate that more straightforward methods can already provide valuable insights. In this paper, we first focus on the attention process and extract causal insights highlighting the crucial roles of some attention heads in this process. Second, we provide an interpretation of the mechanism underlying matching detection.\footnote{Code is available here: \url{https://git.isir.upmc.fr/mat_vast/sigir25-matching-signals}.}
\end{abstract}


\maketitle

\section{Problem statement}

The recent surge in interpretability research stems from a dual motivation: to understand how neural networks accomplish complex tasks and to identify their inherent limitations. In IR, a key research direction is explainability -- i.e., understanding which input components and model elements contribute to predictions. Previous studies have made significant strides in unraveling IR model architectures. For instance, \citeauthor{zhan2020bertanalysis}~\cite{zhan2020bertanalysis} studied cross-encoders and elucidated the role of "[CLS]" and "[SEP]" in the "no-op" operation and demonstrated that relevance prediction is a multi-stage process, involving query and document contextualization, the detection of interaction signals, and finally their composition. Contemporary to this work, \citeauthor{jiang-etal-2021-bert}~\cite{jiang-etal-2021-bert} applies attribution-based method \cite{Schulz2020Restricting} to identify the tokens that contribute the most to the final prediction. Notably, they find that MonoBERT \cite{monobert} still relies on lexical matching but is also able to perform semantic matching, and the "[CLS]" only starts to aggregate relevance signals after layer 16.
Based on these insights, \citeauthor{chen2024axiomatic}~\cite{chen2024axiomatic} further revealed the nuanced roles of specific attention heads in the detection of token duplications and the composition of relevance signals within dual embedding models. These findings suggest a sophisticated, layered information processing mechanism.

Our research extends these investigations by providing a more in-depth analysis of how neural ranking models construct relevance signals. Specifically, we examine the interactions between queries and documents by tracing the information flow through the network. We 1) extract causal insights from the attention process, highlighting the critical roles of specific attention heads in matching, and 2) characterize the nature of these matches. Subsequently, we conduct a deep dive into the internal mechanisms of these matching heads to understand their behavior.

\begin{figure*}[t]
    \centering
    \begin{subfigure}[t]{0.33\linewidth}
        \centering
        \includegraphics[width=\linewidth]{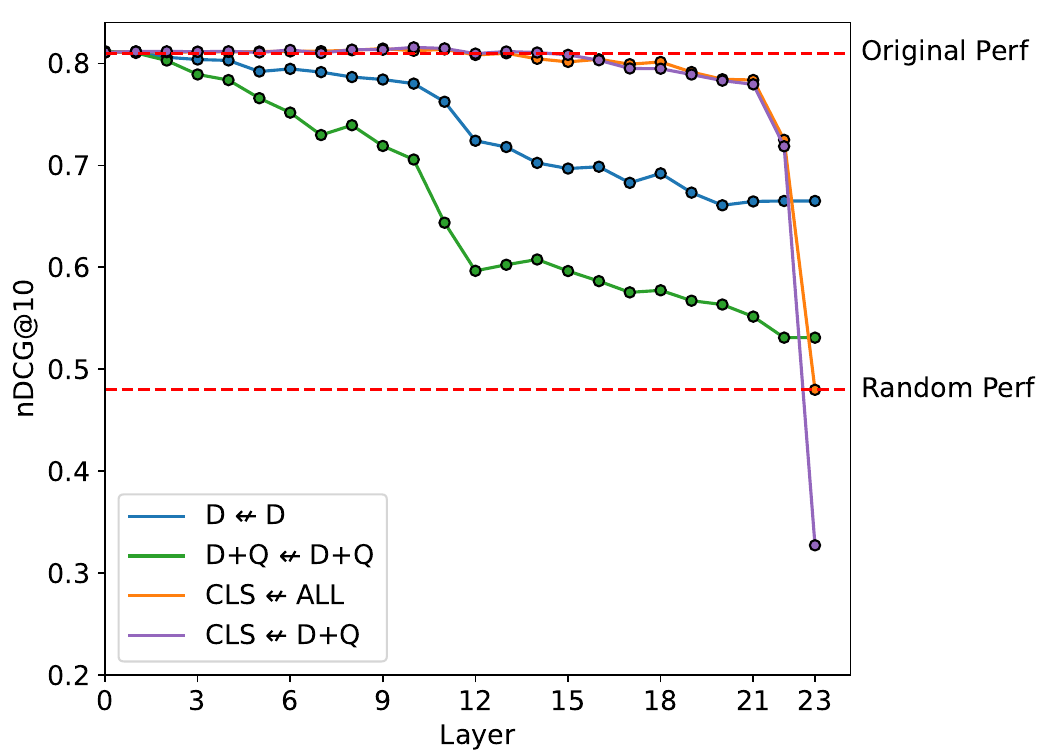}
        \caption{Impact on nDCG@10 of the ablation of a direction up to each layer.}
    \label{fig:ablations}
    \end{subfigure}%
    \begin{subfigure}[t]{0.33\linewidth}
        \centering
        \includegraphics[width=\linewidth]{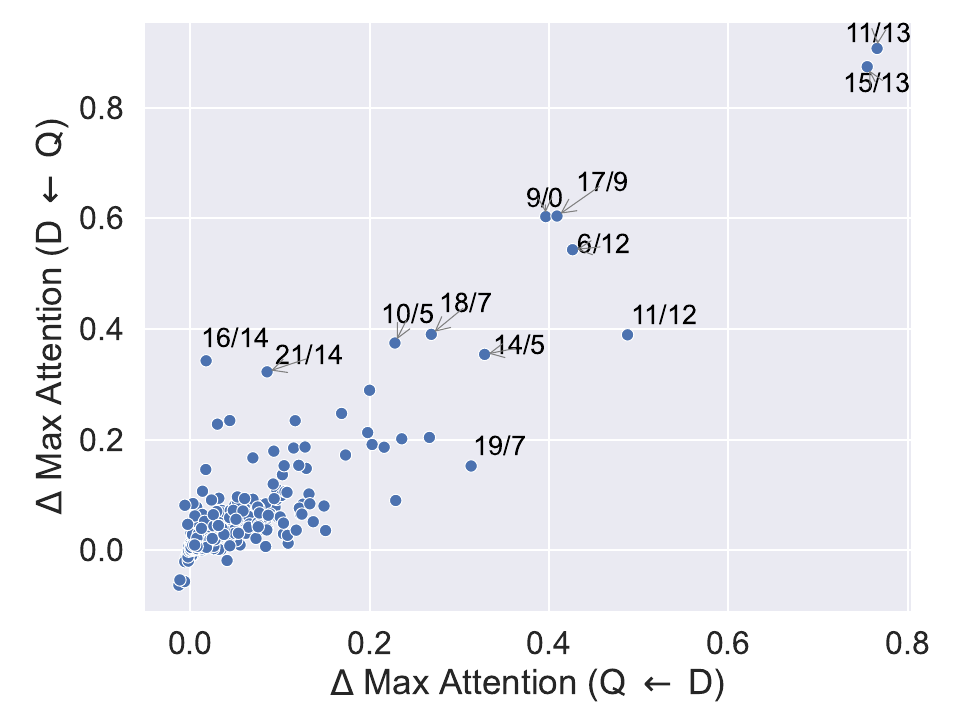}
        \caption{Average diff. in the attention between relevant and easy negatives.}
    \label{fig:scatter_heads}
    \end{subfigure}%
    \begin{subfigure}[t]{0.33\linewidth}
        \centering
        \includegraphics[width=\linewidth]{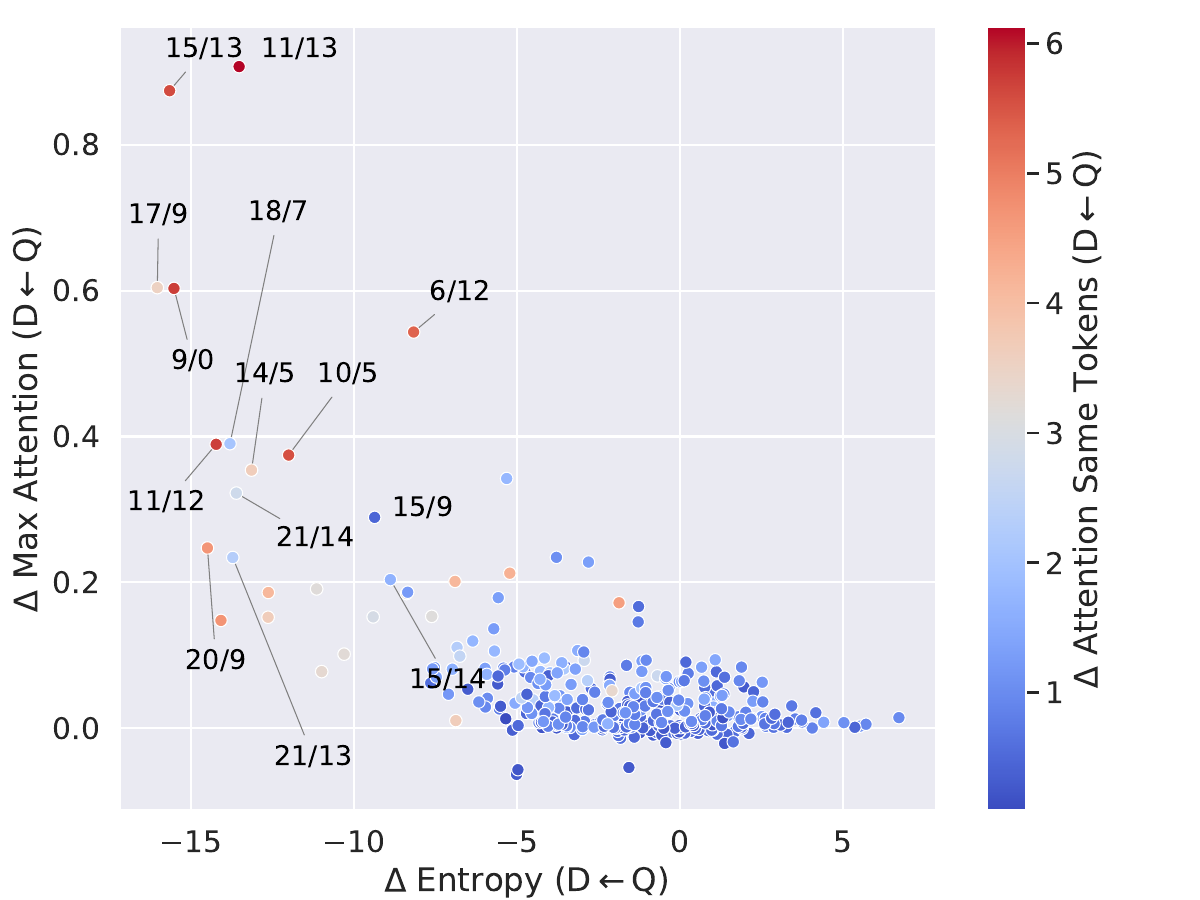}
        \caption{Comparison of attention distributions between relevant and easy negative passages.}
    \label{fig:distribution_attention}
    \end{subfigure}%
    \caption{Summary of the experiments on the attention of Query and Document (layer $\ell$/head $h$).}
\label{fig:summary}
\end{figure*}

\section{Methods}

To do so, we study cross-encoders with an ablation study and the analysis of attention matrices. To gain a better understanding of the transfers of information, we focus on their \emph{directions} across input parts. We split each query-passage pair into 5 distinct parts: the $CLS$ token, the query $Q$, the first separator token $SEP1$, the document $D$, and the last separator token $SEP2$ and look at the attention of the input's part $X$ to the input's part $Y$, a \emph{direction}, within the attention weight matrix or equivalently, the information transmitted by the input's part $Y$ to the input's part $X$ with $X, Y \in \{CLS, Q, SEP1, D, SEP2\}$. \newline
We use the Deep Learning tracks from TREC between 2019 and 2022 \cite{craswell2020overview,craswell2021overview,craswell2022overview,craswell2019overview} to conduct our experiments. The annotations have 4 degrees of relevance $r\in \{0..3\}$. Because documents labeled as "irrelevant" ($r=0$) are still semantically related (a search engine retrieved them), we also study the behavior of MonoBERT in the case where the document is irrelevant \emph{and} not semantically related, by randomly sampling a document in the dataset. We call these documents \textit{easy negatives}, or $r=-1$, as opposed to the \textit{hard negatives}, or $r=0$. This nuance between negatives helps us identify the different types of matching, as we hypothesize that semantic matches are more likely to occur in hard negatives than easy negatives.

\section{Main Results}

\begin{table}[b]
\caption{nDCG@10 results of the ablations. Results in \textbf{bold} indicate a decrease compared to the Original model, and \textbf{*} indicates a decrease with statistical significance ($p\leq0.05$) under the two-tailed Student’s t-test.}
\begin{tabular}{|l|c|c|c|c|c||c|}
\hline
\multicolumn{1}{|c|}{$\nleftarrow$} & CLS & Query & SEP1 & Doc. & SEP2 & {All}\\ \hline
CLS & 0.82 & \textbf{0.79} & 0.81 & 0.81 & 0.81 & \textbf{0.48*}\\ \hline
Query & 0.82 & \textbf{0.78} & 0.81 & \textbf{0.80} & 0.81 & \textbf{0.66*}\\ \hline
SEP1 & 0.81 & 0.81 & 0.81 & 0.81 & 0.81 & 0.81\\ \hline
Doc. & 0.81 & \textbf{0.79} & 0.81 & \textbf{0.67*} & 0.82 & \textbf{0.62*} \\ \hline
SEP2 & 0.81 & 0.81 & 0.81 & 0.81 & 0.81 & 0.81 \\ \hline
\end{tabular}
\caption*{Original perf. = 0.81 / Random ranking = 0.48 (nDCG@10).}
\label{tab:ablations}
\end{table}

With our ablation study, we first validate the importance of the document's contextualization and query-to-document information transfers (Fig. \ref{fig:ablations}). Our results, however, question \citeauthor{zhan2020bertanalysis}~\cite{zhan2020bertanalysis} conclusion on the importance of document-to-query transfers (Table \ref{tab:ablations}). \newline
Second, we study the attention weights between the different parts of the input. We confirm the role in the "no-op" operation of the "[SEP]" and further show that the "[CLS]" might play a similar role through the early to the middle layers. It further unveils attention heads specialized in detecting interactions between queries and relevant passages.
To deepen this observation, we compare the attention weight distributions between documents and queries (Fig. \ref{fig:scatter_heads} and \ref{fig:distribution_attention}) hints at the following behavior:
1) From early to middle layers, these interactions are matching between terms of the query and document (\emph{syntactic matching});
2) From middle layers, matching occurs between semantically-contextualized tokens (\emph{semantic matching}).

\begin{figure}[t]
    \centering
    \includegraphics[width=0.8\columnwidth]{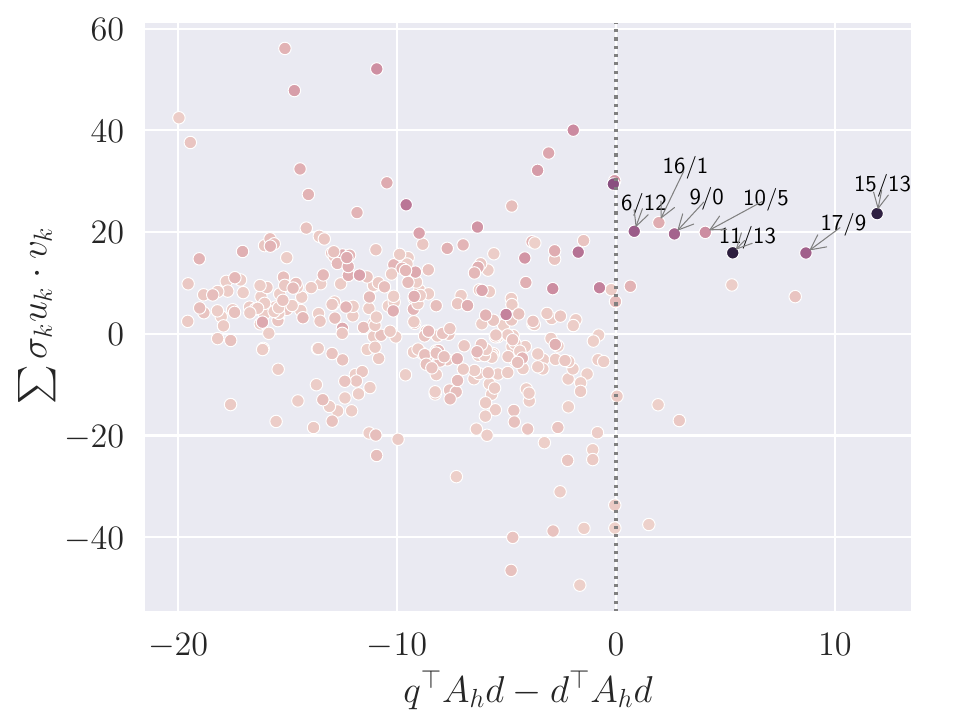}
    \caption{Distribution of the attention heads (layer $\ell$/head $h$) based on their Singular Value Decomposition and their ability to do cross-contextualization. Color levels relate to the average diff. in the attention between relevant and easy negatives (Fig. \ref{fig:scatter_heads}).}
\label{fig:matching_heads}
\end{figure}

Finally, we explain how this matching behavior takes place by analyzing some attention heads $A_h=K_h^\top Q_h$. The logit of the probability that token $i$ attends to $j$ is $x_j^\top A_h x_i$.
 If we suppose that the representation of the $i^{th}$ token of the query and the $j^{th}$ token of the document (at any layer $\ell$ in the model) can be expressed as $E_i = q + x_i + a_i$ and $E_j = d + y_j + b_j$, where $q$ and $d$ respectively represents the part of the embedding of queries and documents shared by every query and document, $x_i$ and $y_j$ their semantic/lexical information, and $a_i$ and $b_j$ some residual information. Then, for a given head $h$, the attention logit for $E_i$ on $E_j$ can be written as:
\begin{align}
    (E_i)^\top A_h E_j &= (q + x_i + a_i)^\top A_h (d + y_j + b_j) \\
    &= q^\top A_h d + x_i A_h y_j + \epsilon
\end{align}

\noindent with $\epsilon \approx 0$.
To validate this hypothesis, we plot the difference between $q^\top A_h d - d^\top A_h d$, which should be greater than 0 when query-document matching is sought, against the propensity of matching tokens with the same meaning, i.e. where $x_i$ and $x_j$ are close. For the latter, we measure to which extent the left and right singular vectors of the Singular Value Decomposition of $A_h$ are similar, i.e. with $A_h=\sum\sigma_ku_kv_k^\top $, we compute $\sum_k \sigma_k u_k \cdot v_k$.
Figure \ref{fig:matching_heads} presents the results. As we can see, the heads identified earlier in Figure~\ref{fig:scatter_heads} tend to follow our hypothesis, i.e. they favor query-document attention rather that document-document or query-query attention, and their left and right singular vectors are more aligned.

\section{Conclusions}

Our work highlights the existence of specific attention heads dedicated to matching across the model's layers as well as the role of the "[CLS]" and "[SEP]" tokens. Further looking at the distribution of the attention weights hints at the following relevance prediction process: 1) Through early to middle layers, query and document are contextualized separately while some attention heads perform lexical matching across them; 2) After tokens have been contextualized, some attention heads start detecting semantic matching signals across query and document. Throughout these two stages, attention heads fallback to the "no-op" operation and attend primarily to the "[SEP]" tokens and to the "[CLS]" token when they can't perform their role ; 3) Finally, relevance information is gathered to the "[CLS]" token representation.
Last, we propose an interpretation of how attention heads detect matching between query and document tokens. Our results hint towards a subspace of their Query-Key matrix dedicated to the detection of query-document matches. 

\begin{acks}
The authors acknowledge the ANR – FRANCE (French National Research Agency) for its financial support of the GUIDANCE project n°ANR-23-IAS1-0003 as well as the CLUSTER IA project n°ANR-23-IACL-0007.
Furthermore, this work was granted access to the HPC resources of IDRIS under the allocation 2024-AD011014444R1 made by GENCI.
\end{acks}

\printbibliography

\end{document}